\documentclass{ws-procs975x65}
\usepackage{graphics}
\begin{document}

\title{The Loop Corrections to the  Parity-Violating  Elastic Electron-Proton Scattering}

\author{Yu Chun Chen$^{*}$}

\address{Institute of Physics, Academia Sinica,\\
Taipei 11529, Taiwan\\
$^*$E-mail: yuchun@phys.sinica.edu.tw}

\author{Chung Wen Kao}

\address{Department of Physics, Chung-Yuan Christian University,\\
Chung-Li 32023, Taiwan\\
E-mail: cwkao@cycu.edu.tw}

\begin{abstract}
We calculate the two-boson-exchange (TBE) corrections to the
parity-violating asymmetry of the elastic electron-proton scattering
in a parton model using the formalism of generalized parton
distributions (GPDs).
\end{abstract}

 \bodymatter

\section{Strangeness content and Parity-Violating measurement}
Strangeness content is one of the most important questions in hadron
structure because it sheds light on the role of the quark-anti quark
sea in the ground state properties. One important tool is elastic
parity-violating electron-proton scattering, which has been used to
probe the the charge and magnetization distributions of the strange
quark within the nucleon. At the tree level, the parity-violating
asymmetry ($A_{PV}$) arises from the interference of diagrams with
one-photon-exchange (OPE) and $Z$-boson exchange shown in Fig 1(a)
and (b), respectively.
\begin{figure}[t]
\centerline{\epsfxsize 1.4 truein\epsfbox{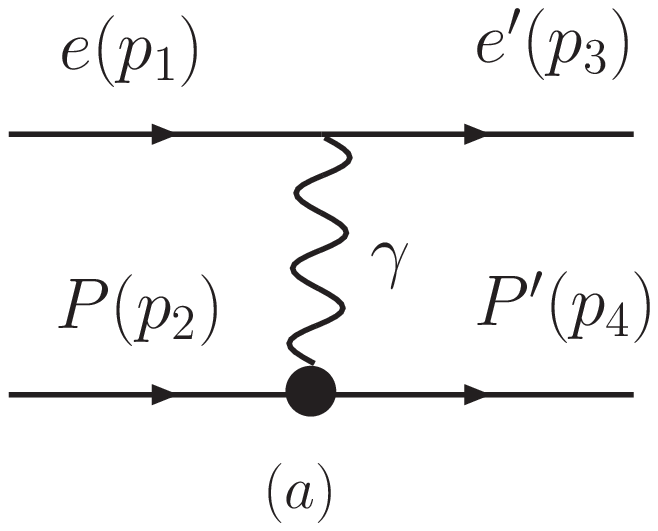} \epsfxsize 1.4
truein\epsfbox{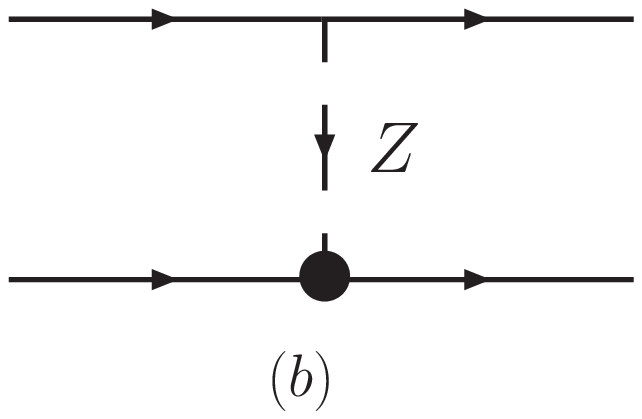}} \centerline{\epsfxsize
1.4truein\epsfbox{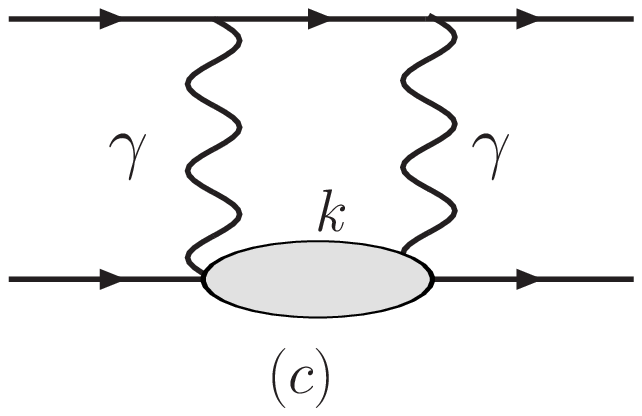} \epsfxsize 1.4 truein\epsfbox{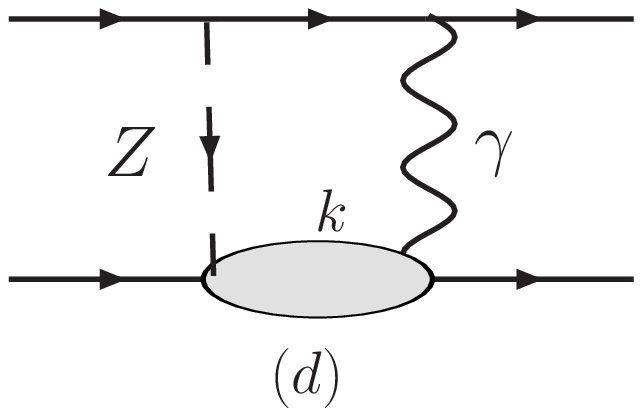}}
\caption{(a) one-photon-exchange, (b) $Z$-boson-exchange, (c) TPE,
and (d) $\gamma Z$-exchange diagrams for elastic {\it ep}
scattering. Corresponding cross-box diagrams are implied.}
\end{figure}
The interactions between photon, $Z$-boson and proton are described
by the form factors of the proton defined as
\begin{eqnarray}
&&\langle p'|J^\gamma_\mu|p\rangle =\overline{u}(p')\left[F^{\gamma
p}_{1}\gamma_\mu+F^{\gamma p}_{2}
\frac{i\sigma_{\mu\nu}q^\nu}{2M}\right]u(p),\nonumber \\
&&\langle p'|J^Z_\mu|p\rangle =\overline{u}(p')\left[F^{Z
p}_{1}\gamma_\mu+F^{Z p}_{2}\frac{i\sigma_{\mu\nu}q^\nu}{2M}
+G^Z_A\gamma_\mu\gamma_5\right]u(p),
\end{eqnarray}
where $q$=$p'-p$ and $M$ is the mass of the proton . $F^{\gamma
p}_{1,2}$ are the form factors of the proton electromagnetic
current, $F^{Z p}_{1,2}$ are the form factors of the proton neutral
weak current. The $A_{PV}$ can be expressed by the form factors
defined above as
\begin{equation}
A_{PV}=\frac{\sigma_R-\sigma_L}{\sigma_R+\sigma_L} = -\frac{G_F
Q^2}{4\pi\alpha\sqrt{2}}\frac{A_E+A_M+A_A}{\bigl[\epsilon (G^{\gamma
p}_{E})^2+ \tau (G^{\gamma p}_{M})^2\bigr]}, \label{A_PV_Born}
\end{equation}
where $G_{F}$ is Fermi constant, $Q^2$=$-q^2$, $\tau$=$Q^2/4M^2$,
and $\epsilon=(1+2 (1+\tau)\tan^2\frac{\theta_e}{2})^{-1}$. $A_{E}$,
$A_{M}$,and $A_{A}$ are defined as $A_E=\epsilon G_{E}^{Z p}
G_{E}^{\gamma p}$, $A_M=\tau G_{M}^{Z p} G_{M}^{\gamma p},$ and
$A_A=(-1+4\sin^2\theta_W)\sqrt{\tau (1+\tau) (1-\epsilon^2)} G_A^Z
G_{M}^{\gamma p}$, where $\theta_W$ is the weak mixing (Weinberg)
angle and $G^{\gamma,Z p}_{E}=F^{\gamma,Z p}_{1}-\tau F^{\gamma,Z
p}_{2}$, $G^{\gamma,Z p}_{M}=F^{\gamma,Z p}_{1}+F^{\gamma,Z p}_{2}$.
When combined with proton and neutron electromagnetic form factors
and with the use of charge symmetry, one obtains the following
relation \cite{Kaplan}:
\begin{equation}
G^{Z p}_{E,M}=(1-4\sin^2\theta_W)G^{\gamma p}_{E,M}-G^{\gamma
n}_{E,M}-G^s_{E,M}. \label{strange}
\end{equation}
$G^{Z p}_{E,M}$ can be extracted from the $A_{PV}$ through
Eq.(\ref{A_PV_Born}) and consequently the strange form factors
$G_{E,M}^{s}$ can be determined through Eq.(\ref{strange}). Four
experimental programs SAMPLE \cite{SAMPLE}, HAPPEX \cite{HAPPEX}, A4
\cite{A4}, and G0 \cite{G0} have been designed to measure the
$A_{PV}$, which is small and ranges from 0.1 to 100 ppm.

The main theoretical uncertainty is from higher-order electroweak
radiative corrections which have been carefully studied in
\cite{Marciano83,Musolf92}. They have been considered the
interference of the one-loop diagrams shown in Figs.~1(c) and 1(d)
in \cite{Marciano83} at both of quark and hadron levels under a zero
momentum-transfer approximation in the low energy limit. To refine
our knowledge it is important to go beyond the ``zero
momentum-transfer approximation''.
\begin{figure}
\begin{center}
\includegraphics[width=5.0cm]{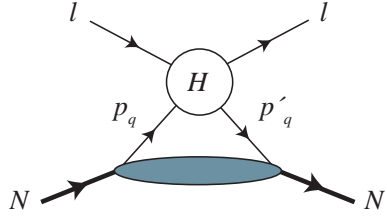}
\caption{Handbag approximation for the elastic lepton-nucleon
scattering. In the partonic process indicated by H, the lepton
scatters from quarks within the nucleon, with momenta $P_{q}$ and
$P_{q'}$. The lower blob represents the GPD's of the nucleon.}
\label{handbag}
\end{center}
\end{figure}

At high $Q^2$, the two-photon-exchange (TPE) box diagram was
suggested \cite{Guichon03} to explain the discrepancy between the
measurement of the proton electric to magnetic form factor ratio
$R=G_E/G_M$, from Rosenbluth technique and polarization transfer
technique\cite{Jone00}. The TPE effect has been evaluated in
\cite{Chen04} in a parton model using GPD's where the handbag
approximation is used. The contribution of the interference of the
TPE process of Fig.~1(c) with diagram of Figs.~1(a) and 1(b) to the
$A_{PV}$, has been evaluated in \cite{Afanasev05} in the same parton
model used in \cite{Chen04}. It was found that indeed the TPE
correction to the $A_{PV}$ can reach several percent in certain
kinematics, becoming comparable in size with existing experimental
measurements of strange-quark effects in the proton neutral weak
current.

On the other hand, the contribution of the interference of the TPE
process of Fig.~1(c) with diagram of Figs.~1(a) and 1(b) to the
$A_{PV}$, and the interference of the $\gamma$-$Z$ exchange process
of Fig.~1(d) with diagram of Fig.~1(a) have both been evaluated in
\cite{Zhou07} in a hadronic model where the excited intermediate
states are neglected and the on-shell nucleon form factors are
inserted. It has been found that these effects own strong $\epsilon$
and $Q^2$ dependencies in the parity-violating $ep$ scattering.

\section{Formalism with parionic calculation}
In this Letter, we first calculate the TPE and $\gamma Z$ exchange
corrections on a quark $l(k)+q(P_{q})\rightarrow l(k')+q(P_{q'})$ ,
denoted by the scattering amplitude $H$ in Fig.~(\ref{handbag}).
Subsequently we embed the quarks in the proton as described through
the nucleon's GPD's. Here only GPD's of $u$ and $d$ quarks are
included because the strange quark contribution to the box diagrams
are very small.

\begin{figure}
\begin{center}
\includegraphics[width=10.0cm]{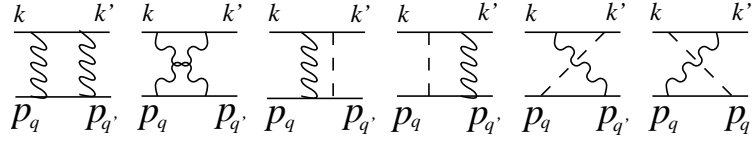}
\caption{Direct and crossed box diagrams to describe the two-photon
exchange and $\gamma$-$Z$ contribution to the lepton-quark
scattering processes.} \label{direct}
\end{center}
\end{figure}

In the standard model, the electromagnetic current and the neutral
weak current of quarks are $J_{\mu}^{em}=\sum_{f=u,d} Q_{f}
\bar{q}_{f}\gamma_{\mu}q_{f},
J_{\mu}^{Z}=\sum_{f=u,d}\bar{q}_{f}(g^{f}_{V}+g^{f}_{A}\gamma_{5})q_{f},$
here we follow the notation of \cite{Musolf92}. The quark-level
parity-violating amplitudes of the $\gamma$-$Z$ exchange box
diagrams in Fig.~\ref{direct} are
\begin{equation}
{\cal M}^{PV}_{\gamma Z}=\frac{-iG_{F}}{2\sqrt{2}}\sum_{q=u,d}
\left[t^{1}_{q}\bar{u}_{e}\gamma_{\mu}\gamma_{5}u_{e}\bar{u}_{q}^{\mu}u_{q}
+t^{2}_{q}\bar{u}_{e}\gamma_{\mu}u_{e}\bar{u}_{q}^{\mu}\gamma_{5}u_{q}\right],
\end{equation}
where
\begin{eqnarray}
\frac{t^{q}_{1}}{M_{Z}^{2}}&=&e^2e_{q}\left[c_{1}^{hard}g^{e}_{A}g^{q}_{V}+c_{2}g^{q}_{A}g^{e}_{V}\right],\nonumber \\
\frac{t^{q}_{2}}{M_{Z}^{2}}&=&e^2e_{q}\left[c_{1}^{hard}g^{e}_{V}g^{q}_{A}+c_{2}g^{q}_{V}g^{e}_{A}\right],
\end{eqnarray}
with $g^{e}_{V}$=$-1+4\sin^{2}\theta_{W}$, $g^{e}_{A}$=1 and
$c_{1}^{hard}$ and $c_{2}$ are defined as
\begin{eqnarray}
&&c_{1}=\frac{-1}{4\pi^2
M_{Z}^{2}}\left[\ln\left(\frac{\lambda^2}{Q^2}\right)+\frac{\pi^2}{2}\right]+
\frac{3}{16\pi^2 M_{Z}^{2}}\ln\left(\frac{\hat{u}}{\hat{s}}\right),\nonumber \\
&&c_{2}=\frac{1}{16\pi^2
M_{Z}^{2}}\left[-7+3\ln\left(\frac{\hat{s}}{M_{Z}^{2}}\right)+3\ln\left(\frac{\hat{u}}{M_{Z}^{2}}\right)\right].
\end{eqnarray}
$\hat{s}=(P_{q}+k)^2,\hat{u}=(P_q-k')^2$ and $\lambda$ is the
infrared cut-off input by infinitesimal photon mass. The amplitudes
are separated into the soft and hard parts, the soft part
corresponds with the situation where the photon carries zero four
momentum and one obtains $c_{1}^{soft}=\frac{-1}{4\pi^2
M_{Z}^{2}}\left[\ln\left(\frac{\lambda^2}{Q^2}\right)+\frac{\pi^2}{2}\right]$,
therefore the hard part is $c_{1}^{hard}=\frac{3}{16\pi^2
M_{Z}^{2}}\ln\left(\frac{\hat{u}}{\hat{s}}\right)$. The IR
divergence arising from the direct and crossed box diagrams is
concealed with the bremsstrahlung interference contribution with a
soft photon emitted from the electron and the proton. We limit
ourselves to discuss the hard $\gamma$-$Z$ exchange contribution in
this article.

The next step is to calculate the $ep$ amplitudes. These amplitudes
are obtained as a convolution between an electron-quark hard
scattering and a soft nucleon matrix element. The procedure is
similar to the one in \cite{Chen04}. In the kinematics regime where
$s$=$(p+k)^2$, $-u$=$-(k-p')^2$ and $Q^2$ are all large enough
compared to hadronic scale $\Lambda_{H}$ which we choose as 1GeV.
The parity-violating $T$-matrix can be written as
\begin{eqnarray}
&&T^{PV,hard}_{h,\lambda'_{N}\lambda_{N}}=-i\frac{G_{F}}{2\sqrt{2}}\left\{\bar{u}(k',h)\gamma_{\mu}u(k,h)
\bar{u}_{N}(p',\lambda'_{N})\right.\nonumber \\
&&\left.[\delta
G_{A}^{PV}\gamma^{\mu}]\gamma_{5}u_{N}(p,\lambda_{N})
+\bar{u}(k',h)\gamma_{\mu}\gamma_{5}u(k,h)
\bar{u}_{N}(p',\lambda'_{N})\right.\nonumber \\
&&\left.\left[\delta G_{M}^{PV}\gamma^{\mu}-\delta
F_{2}^{PV}\frac{P^{\mu}}{M}\right]u_{N}(p,\lambda_{N}) \right\}.
\end{eqnarray}
where
\begin{eqnarray}
\delta G_{M}^{PV}&=&\frac{1+\epsilon}{2\epsilon}D-\frac{1+\epsilon}{2\epsilon}\frac{Q^2+4M^2}{s-u}F, \nonumber \\
\delta F_{2}^{PV}&=& \frac{1}{1+\tau}\left[\delta G^{PV}_{M}-\sqrt{\frac{1+\epsilon}{2\epsilon}}E\right], \nonumber \\
\delta G^{PV}_{A}&=&
\frac{1+\epsilon}{2\epsilon}F-\frac{1+\epsilon}{2\epsilon}\frac{Q^2}{s-u}D,
\end{eqnarray}
with
\begin{eqnarray}
D&\equiv&\int^{1}_{-1}\sum_{q=u,d}\frac{dx}{x}\frac{Q^2 t^{q}_{1}+(\hat{s}-\hat{u})t^{q}_{2}}{s-u}(H^{q}+E^{q}),\nonumber \\
E&\equiv&\int^{1}_{-1}\sum_{q=u,d}\frac{dx}{x}\frac{Q^2
t^{q}_{1}+(\hat{s}-\hat{u})t^{q}_{2}}
{s-u}(H^{q}-\tau E^{q}),\nonumber \\
F&\equiv&\int^{1}_{-1}\sum_{q=u,d}\frac{dx}{x}\frac{Q^2
t^{q}_{2}+(\hat{s}-\hat{u})t^{q}_{1}} {s-u}sgm(x)\tilde{H}^{q},
\label{DEF}
\end{eqnarray}
here $H^{q}$, $E^{q}$ and $\tilde{H}^{q}$ are the GPD's for a quark
in the nucleon. To estimate the amplitudes of Eq. (\ref{DEF}) one
needs to specify a model for the GPD's. Again we follow
\cite{Chen04} to adopt an unfactorized model of GPD's in terms of a
forward parton distributions and a gaussian factor in $x$ and $Q^2$.

\section{Results}
\begin{figure}[t]
\centerline{\epsfxsize 1.4 truein\epsfbox{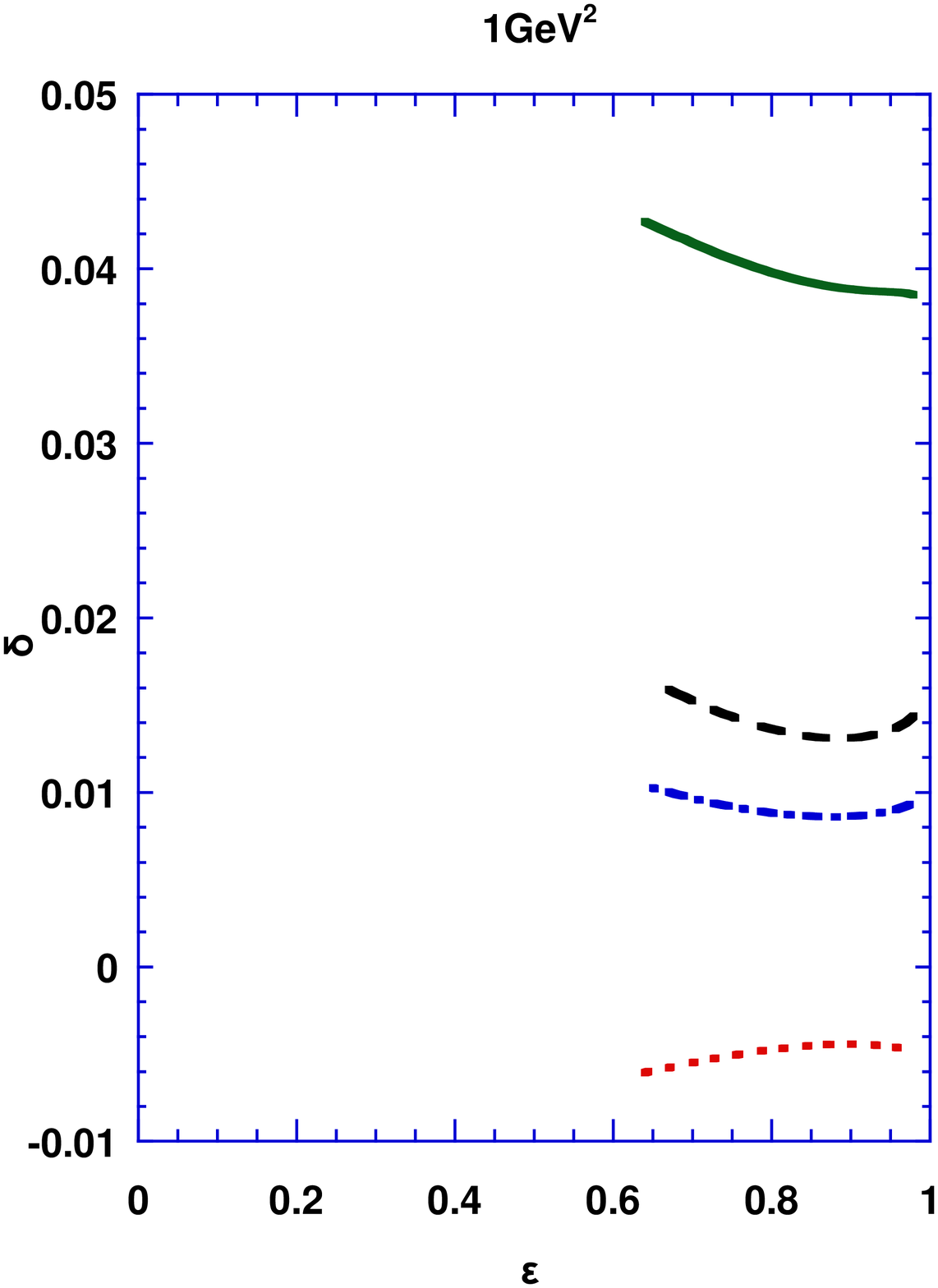} \epsfxsize
1.4 truein\epsfbox{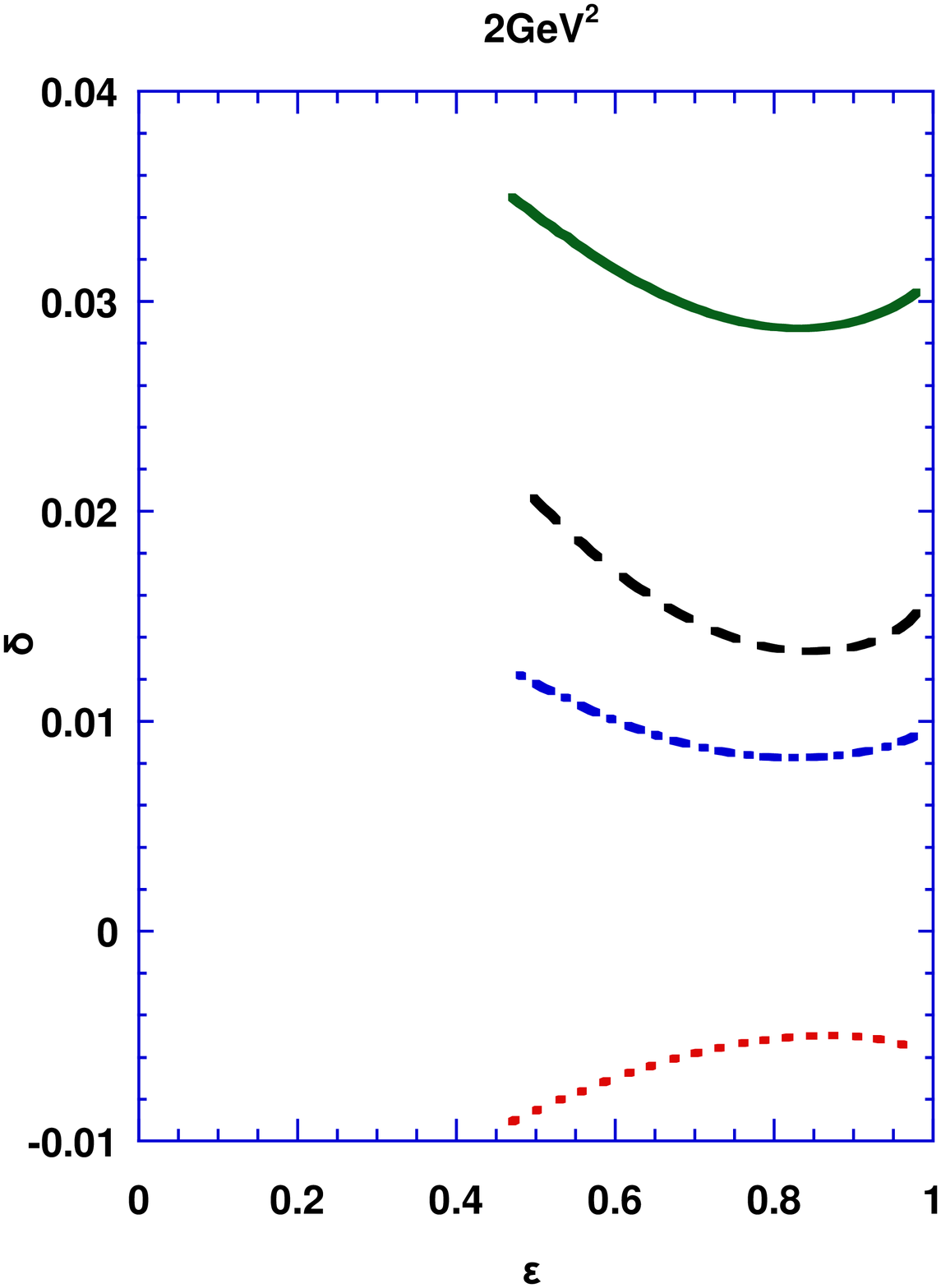}} \centerline{\epsfxsize
1.4truein\epsfbox{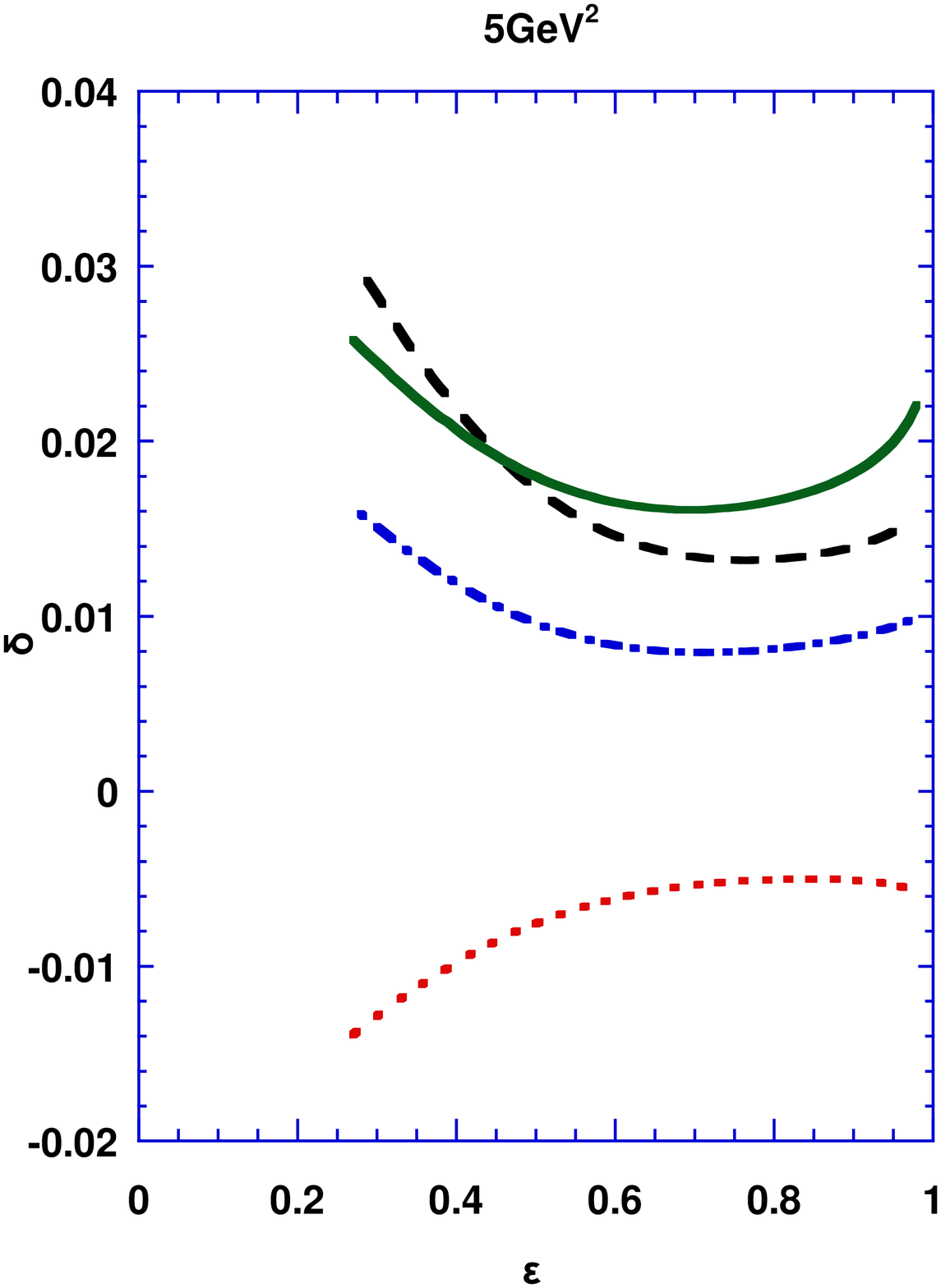} \epsfxsize 1.4
truein\epsfbox{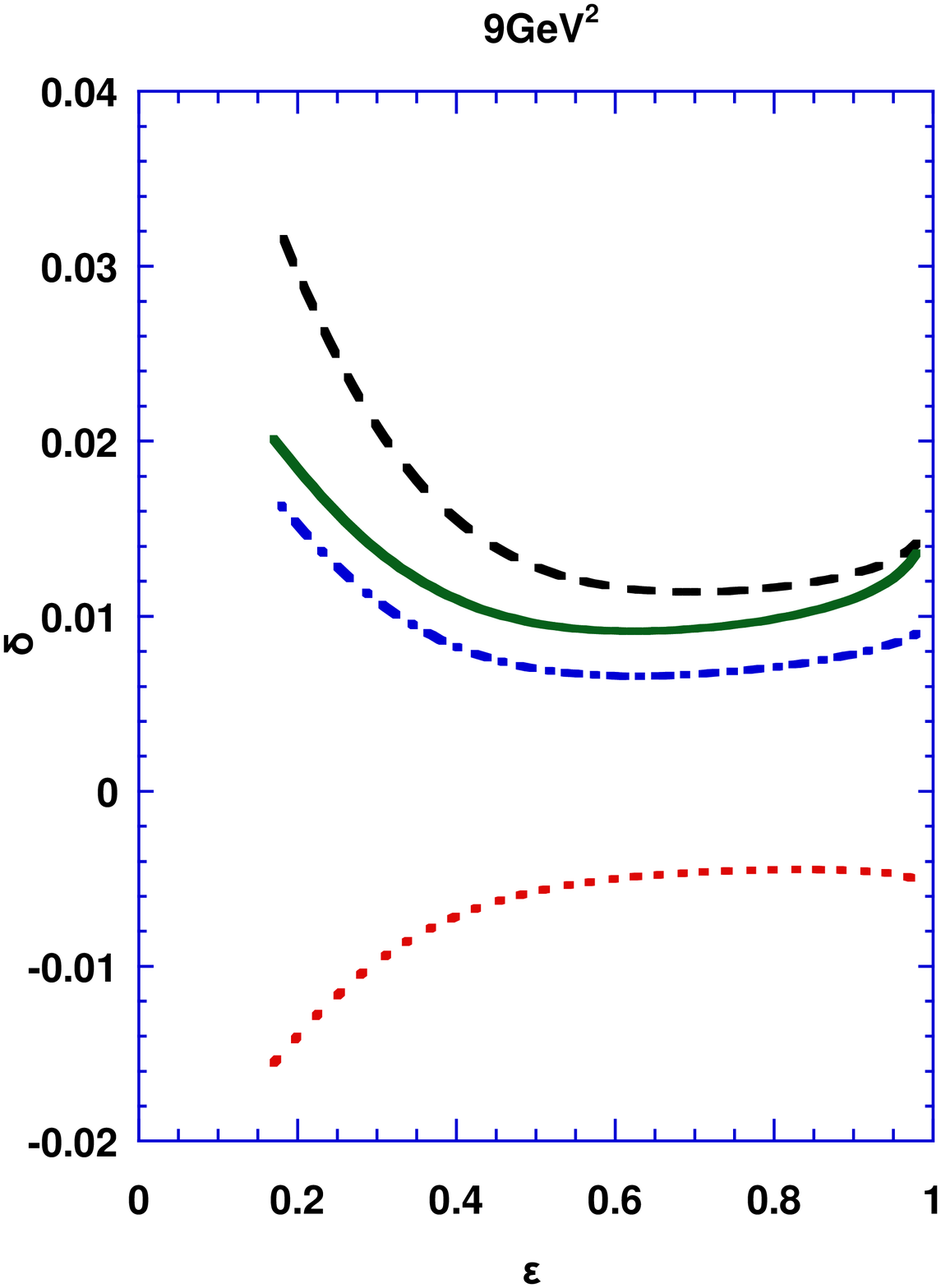}}
 \caption{TPE and $\gamma$-Z exchange
corrections to the $A_{PV}$ as functions of $\epsilon$. The left of
upper panel for $Q^2$=1.0 GeV$^2$, the right for $Q^2$=2.0 GeV$^2$.
The left of lower panel for $Q^2$=5.0GeV$^2$, the right for
$Q^2$=9.0GeV$^2$. $1\gamma\times2\gamma$ (dot line) denotes to the
correction only coming from the interference of FIG.1(a) and
FIG.1(c), $Z\times2\gamma$ (dash line) denotes to the correction
only coming from interference of FIG.1(b) and FIG.1(c), $2\gamma$
(dash dot line) denotes to $1\gamma\times2\gamma$+$Z\times2\gamma$,
Total (solid line) denotes to the full TPE and $\gamma$-Z exchange
corrections including the interference of FIG.1(a) and FIG.1(d)}
\label{Mainresult}
\end{figure}

In Fig. \ref{Mainresult}, we show the TPE and $\gamma Z$-exchange
corrections to the $A_{PV}$ by plotting $\delta$, defined by
\begin{equation} A_{PV}(1\gamma+Z+2\gamma+\gamma
Z)=A_{PV}(1\gamma+Z)(1+\delta),\end{equation} at four different
values of $Q^2= 1.0, \,2.0,\, 5.0$ and $9.0 \,$ GeV$^2$.
$A_{PV}(1\gamma+Z)$ denotes the $A_{PV}$ arising from the
interference between OPE and $Z$-boson-exchange, i.e., Figs. 1(a)
and 1(b) while $A_{PV}(1\gamma+Z+2\gamma+\gamma Z)$ includes the
effects of TPE and $\gamma Z$-exchange. The full results are
represented by solid curves. We also show in Fig. 2 the
interferences between OPE and TPE ($1\gamma\times2\gamma$), by
dotted lines, as well as that between $Z$-exchange and TPE
($Z\times2\gamma$) in dashed lines, with their sum ($2\gamma$)
denoted by dot-dashed lines.

We first reproduce the result in \cite{Afanasev05}, which
demonstrates that $1\gamma\times2\gamma$ contribution partially
cancels that of $Z\times2\gamma$ and their sum are always positive.
The similar feature also appears in the result of hadronic model
calculation \cite{Zhou07}. However, in \cite{Zhou07} those curves
all decrease when $\epsilon$ increase and vanish when $\epsilon$=1
but our GPD's calculation results show different feature for that
and with much higher value. It needs further study to understand the
reason of this difference. Furthermore, the solid lines and the
dot-dashed lines are found to be almost parallel. It means that
their difference, namely the $\gamma$-$Z$ effects, essentially own
very mild $\epsilon$ dependencies.

\begin{figure}
\begin{center}
\psfig{file=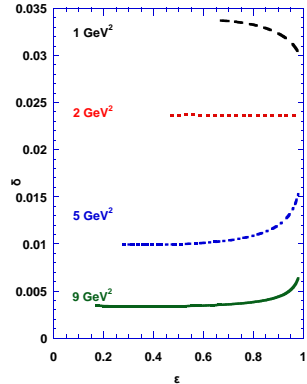, width=4.0cm} \caption{$ZZ$ and
$WW$-exchange diagrams contribution to the $A_{PV}$ for elastic {\it
ep} scattering.} \label{aba:ZZ}
\end{center}
\end{figure}

There are still some box diagrams which are not included in the
hadronic model calculation. Those are the diagrams with two-Z-boson
exchange (TZE) diagrams and two-W-boson exchange (TWE) exchange. We
also calculate these diagrams by using same parton model. In the
hadronic model their contributions are expected to be very small,
because of the main contribution of loop integral is from low loop
momentum due to the inserted form factors functioning as regulators.
However, in the partonic model, the main contribution is from high
loop momentum and it is straightforward to calculate it in this
letter. The sum of TZE and TWE effect in our model is demonstrated
at Fig.~\ref{aba:ZZ}. It is clear the those effect varies with $Q^2$
in the similar way as other effects, namely it increases when $Q^2$
decrease. Besides, we find that effect becomes sensitive to
$\epsilon$ in high $\epsilon$ region.

\section{Conclusions}
We conclude that the TBE exchange effects contribute to $A_{PV}$ at
a level of one or more per cent. Most of them are very sensitive to
$Q^2$ and their magnitudes increase when $Q^2$ decrease, except TPE
exchange effects. Hence, the claim made by \cite{Zhou07} that
$\gamma$-$Z$ exchange effect is overestimated in the zero momentum
transfer approximation is also supported by partonic calculation.
Besides, we find that two-Z/W-boson-effect is not negligible in the
partonic calculation and it is about same magnitude as TPE and
$\gamma$-$Z$ exchange effect. These TBE effects are very important
when extracting the s-content of nucleon, but the impact of these
diagrams to extract strangeness is in the further study.

\section{Acknowledgements}
This work is partially supported by the National Science Council of
Taiwan under grants nos. NSC95-2112-M033-003-MY3 (C.W.K.) and
NSC96-2811-M-001-087 (Y.C.C). The authors would like to thank M.
Vanderhaeghen for his helpful opinions for this study.

\bibliographystyle{ws-procs975x65}
\bibliography{ws-pro-sample}

\begin{thebibliography}{99}
\bibitem{Kaplan} D. Kaplan, A. Manohar, Nucl. Phys. {\bf B 310},
                 527 (1988).
\bibitem{SAMPLE} B. Mueller {\it et al.}, Phys. Rev. Lett. {\bf 78},
                 3824 (1997); D.~T. Spayde {\it et al.}, Phys. Lett.
                 {\bf B 583}, 79 (2004).
\bibitem{HAPPEX} K.~A. Aniol {\it et al.} (HAPPEX), Phys. Rev. {\bf C 69},
                 065501 (2004), Phys. Lett. {\bf B 635}, 275 (2006);
                 A. Acha {\it et al.} (HAPPEX), Phys. Rev. Lett. {\bf 98},
                 032301 (2007).
\bibitem{A4}     F.~E. Maas {\it et al.} (A4), Phys. Rev. Lett. {\bf 93},
                 022002 (2004); Phys. Rev. Lett. {\bf 94}, 152001 (2005);
                 B. Glaser (for the A4 collaboration) Eur. Phys. J.
                 {\bf A 24, S2}, 141(2005).
\bibitem{G0}     D.~S. Armstrong {\it et al.} (G0), Phys. Rev. Lett.
                 {\bf 95}, 092001 (2005); C. Furget (for the G0 collaboration),
                 Nucl. Phys. Proc. Suppl. {\bf 159}, 121 (2006).
\bibitem{Marciano83} W.~J. Marciano and A. Sirlin, Phys. Rev. {\bf D 27},
                 27 (1983);Phys. Rev. {\bf D 29}, 75 (1984).
\bibitem{Musolf92} M.~J. Musolf and T.~W. Donnelly, Nucl. Phys. {\bf A 546},
                 509 (1992); M.~J. Musolf, {\it et al.}, Phys. Rep.
                 {\bf 239} 1 (1994), Phys. Rev. {\bf C 60}, 015501 (1999).
\bibitem{Guichon03} P.~A.~M. Guichon and M. Vanderhaeghen, Phys. Rev. Lett.
                 {\bf 91}, 142303 (2003).
\bibitem{Jone00} M.~K. Jones {\it et al.}, Phys. Rev. Lett. {\bf 84},
                 1398 (2000); O. Gayou {\it et al.}, Phys. Rev.
                 Lett. {\bf 88}, 092301 (2002).
\bibitem{Chen04} Y.~C. Chen, A.~V. Afanasev, S.~J. Brodsky, C.~E. Carlson,
                 M. Vanderhaeghen, Phys. Rev. Lett {\bf 93}, 122301 (2004).
\bibitem{Afanasev05} A.~V. Afanasev, C.~E. Carlson, Phys. Rev. Lett. {\bf 94},
                 212301 (2005).
\bibitem{Zhou07} H.~Q. Zhou, C.~W, Kao and S.~N. Yang, Phys. Rev. Lett
                 {\bf 99}, 262001 (2007).

\end{thebibliography}

\end{document}